\documentclass[journal=jpccck,manuscript=article]{achemso}

\usepackage[version=3]{mhchem}
\usepackage{hyperref}
\usepackage{xcolor}

\newcommand{\revadd}[1]{#1}

\author{Nicolas Castel}
\affiliation{Chimie ParisTech, PSL Research University, CNRS, Institut de Recherche de Chimie Paris, 75005 Paris, France}
\alsoaffiliation{\'{E}cole des Ponts, 77420 Marne-la-Vall\'{e}e, France}
\author{Fran\c{c}ois-Xavier Coudert}
\email{fx.coudert@chimieparistech.psl.eu}
\affiliation{Chimie ParisTech, PSL Research University, CNRS, Institut de Recherche de Chimie Paris, 75005 Paris, France}
\title{Atomistic Models of Amorphous Metal--Organic Frameworks}

\begin{document}

\begin{abstract}
There is an increasing interest in the amorphous states of metal-organic frameworks (MOFs) and porous coordination polymers, which can be produced by pressure-induced amorphization, temperature-induced amorphization, melt-quenching, ball milling, irradiation, etc. They can exhibit useful physical and chemical properties, distinct from those achievable in the crystalline states, along with greater ease of processing, and intrinsic advantages over crystals and powders, such as high transparency and mechanical robustness. However, these amorphous states are particularly challenging to characterize, and the determination of their framework structure at the microscopic scale is difficult, with only indirect structural information available from diffraction experiments. In this Perspective, we review and compare the existing methodologies available for the determination of microscopic models of amorphous MOFs, based on both experimental data and simulation methods. In particular, \revadd{we present} the atomistic models that can be obtained using Reverse Monte Carlo (RMC) methods, Continuous Random Networks (CRN), classical and \emph{ab initio} molecular dynamics, reactive force fields, and simulated assembly/polymerization algorithms.
\end{abstract}


\vspace{2cm}

\section{Introduction}

Metal-organic frameworks (MOFs) have been proposed for a large variety of applications, due to their hybrid organic--inorganic nature, their porosity, and the tunability of both their chemical composition and framework. Of the rapidly increasing number of studies of MOFs in the research literature, most are focused on perfectly ordered, crystalline structures. The vast majority of reported MOF structures are crystalline, because of ease of experimental identification and characterization with common laboratory equipment. However, the dynamic nature of MOFs is a key characteristic of these relatively weakly bonded frameworks, compared to more traditional porous solids like zeolites --- and accounts in a large part for their appeal, and some of their extraordinary chemical and physical properties. Throughout this family of materials, many authors have noted the common occurrence of large-scale flexibility under stimulation, the presence of crystallographic defects, and the possibility of correlated disorder.\cite{Bennett2021}

However, the definition of MOFs does not restrict this term to crystalline phases\cite{Batten2013} --- although the reliance on crystalline databases in counting the ``number of known MOF materials'' offers an example of clear bias towards crystals. Recently, the number of noncrystalline MOF states (and porous coordination polymers) reported have seen a rapid expansion.\cite{Bennett2018} These include the recently discovered MOF liquids, MOF gels (forming aerogels or monoliths upon drying), and a large variety of glassy states and amorphous solids, which can be produced by pressure-induced amorphization, temperature-induced amorphization, melt-quenching, ball milling, irradiation, etc. These non-crystalline states possess useful physical and chemical properties \revadd{distinct from those achievable in the crystalline phases, such as isotropy, the absence of grain boundaries, high transparency, and mechanical robustness, while retaining intrinsic advantages of crystals and powders.} They also allow for a greater ease of processing and have been proposed for several industrial applications. \cite{Bennett2014, Fonseca2021}

In balance with these promising properties, amorphous states are particularly challenging to characterize and their framework structures at the microscopic scale hard to determine. Indirect structural information can be available from diffraction experiments, but unlike for crystals, it cannot be solved into a nice periodic atomic structure as a matter of routine analysis. Moreover, the computational description of disordered materials is also more complex than that of crystals\revadd{, firstly due to their lack of periodicity and symmetry. As it is inherently impossible to reduce an amorphous material to a small finite set of atomic coordinates, each atomic structure should best be seen as representative of a plausible local configuration. \cite{Massobrio2015} To complicate matters further, the structure and properties of many amorphous systems -- including most glasses -- depend on their formation route, calling for specific methodological developments of plausible numerical analogues to experimental production processes. \cite{Li2017}  It is all the more difficult when} the formation of the amorphous phase involves changes \revadd{in coordination which severely limits the use of every method relying on a classical description of the interactions \cite{Gaillac2020}.}

For these reasons, there are only very few atomistic models of amorphous MOFs \revadd{-- mostly glasses --} available in the literature, from both experimental and computational studies. In this Perspective, we review and compare the existing methodologies available for the determination of microscopic models of amorphous MOFs, as well as the structure they generated, assessing their strengths and weaknesses.

\section{Reverse Monte Carlo (RMC) models}
The determination of material structures from limited, indirect experimental data, is a classic example of an inverse problem. Reverse Monte Carlo (RMC) modeling is a general method for the resolution of such inverse problems, which in condensed matter aims to produce atomistic models based on available experimental data, in particular from X-ray or neutron-scattering experiments.\cite{McGreevy1988} Starting from an initial configuration provided by the user, atomic positions are iteratively adjusted using a Monte Carlo algorithm to minimize the difference between calculated and experimental total structure factors. The minimization can be performed under a set of constraints (density, coordination, bond lengths or angles, etc.) which are added to the structure factor difference with customizable weights. No knowledge of the physics or chemistry of the system is required; for example, no interatomic potential is used in the conventional sense. \cite{McGreevy2001} This makes RMC modeling applicable to any system, a very appealing property for new classes of materials. \cite{Playford2014} The downside is that the thermodynamical consistence of the model is not guaranteed, as the optimized configuration could be physically unrealistic. The problem is also typically underconstrained, as a large number of different configurations could equally fit the experimental data.

\begin{figure}[ht]
\centering
\includegraphics[width=1.0 \textwidth]{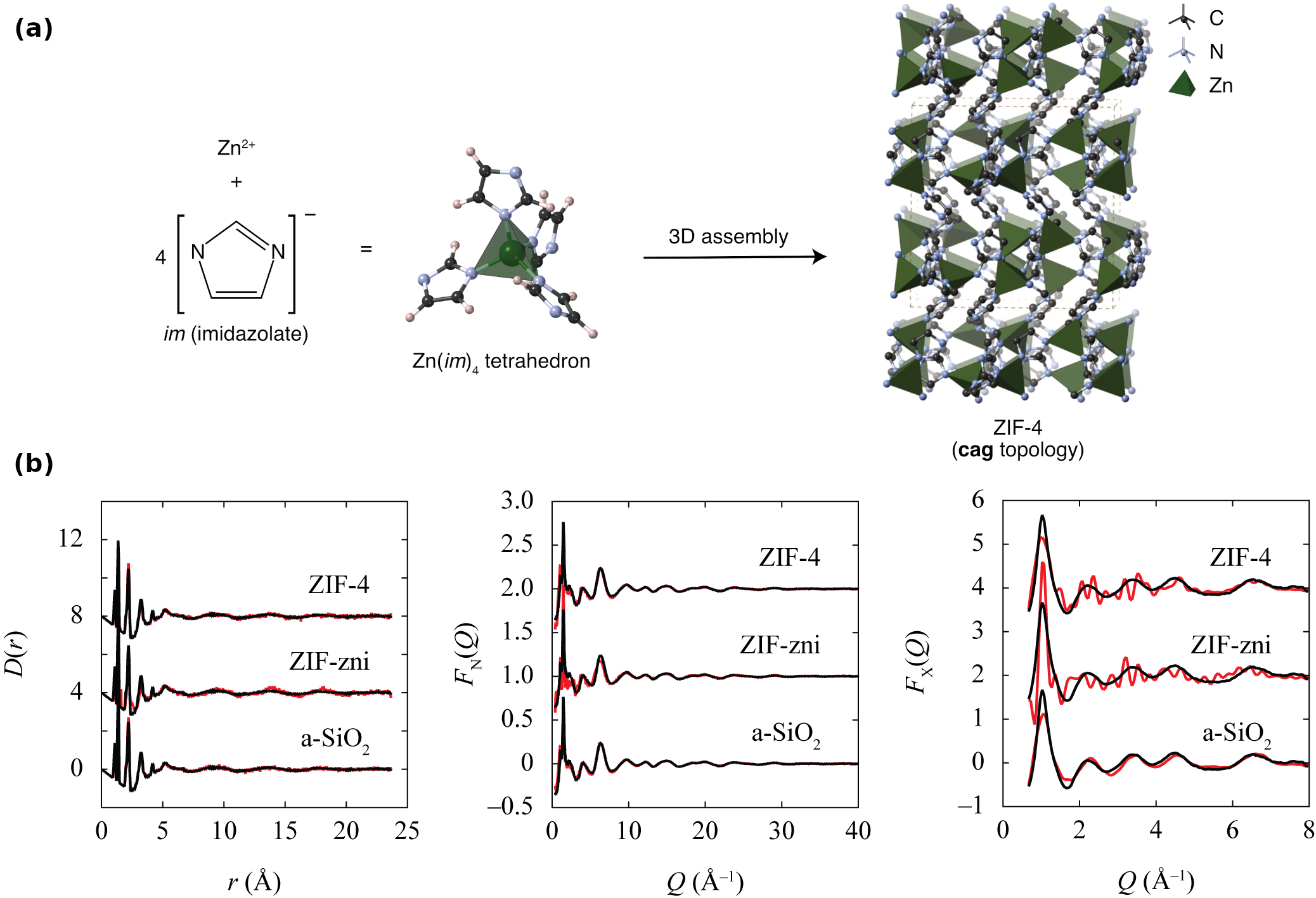}
\caption{
(a) Representation of the assembly of ZIF-4 as a three-dimensional network of Zn(Im)$_4$ tetrahedra.
(b) RMC fits (red) to experimental data (black) of the differential pair correlation function $D(r)$, neutron and X-ray total scattering functions $F_N(Q)$ and $F_X(Q)$. Calculated with three initial configurations: ZIF-4 (top), ZIF-zni (middle) and an $a$-\ce{SiO_2} CRN (bottom). 
Reprinted with permission from Ref.~\citenum{Bennett2010}. Copyright 2010 by the American Physical Society.}
\label{fig:rmc}
\end{figure}

For these reasons, RMC modeling was used to generate the first published model of an amorphous MOF, an amorphous ZIF-4 ($a$-ZIF-4) obtained by temperature-induced amorphization (TIA)\revadd{\cite{Bennett2010, Beake2013}} --- and is still, to date, the only technique that has been used to construct an amorphous MOF model from experimental data. 
ZIF-4 is a zeolitic imidazolate framework (ZIF) composed of Zn$^{2+}$ metal nodes and imidazolate (Im) organic linkers organized as \ce{Zn(Im)_4} tetrahedra linked by Zn--N coordinative bonds as illustrated on \autoref{fig:rmc}a.
This $a$-ZIF-4 model was constructed from neutron and X-ray total scattering data collected at a synchrotron facility. In addition to this experimental data, it used density, connectivity and molecular geometry constraints to preserve the network topology of the initial configuration. Three initial configurations were tested: two crystalline polymorphs (ZIF-4 and ZIF-zni), and a Continuous Random Network (CRN) model of $a$-\ce{SiO_2} adapted by substituting atom groups (see next section and \autoref{fig:crn_substitution}c for details). As shown on \autoref{fig:rmc}b, only the latter configuration allowed RMC refinements to capture the experimental data and resolve its structure, which was therefore confirmed to be highly disordered and not crystal-like. Additionally, the very process of RMC modeling with different initial configurations showed that the system undergoes a reconstructive \revadd{(i.e. involving bond breaking and forming events)} phase transition during amorphization, and hints that the framework topology is markedly different from the crystal phases.

Later works by Keen and Bennett on different ZIF systems have adopted the same method and improved it by inputting experimental values of the density, as obtained from pycnometric density measurements. It was first extended to produce a model of ZIF-8 (featuring 2-methylimidazolate ligands instead of imidazolate) amorphized by ball milling \cite{Cao2012}, starting from the same CRN model with a change of organic ligand. A reconstructive phase transition was also demonstrated. Then it was applied to amorphous phases of ZIF-4 produced by two other routes, which were modeled and compared to temperature induced amorphization: melt--quenching \cite{Gaillac2017} and ball milling. \cite{Keen2018} The atomistic models were found to be near-identical, with indistinguishable short-range order. To have more insight into the structural changes during the melt-quench\revadd{ing} procedure, an intermediate model of the liquid state was also fitted to X-ray data, although the accuracy of RMC for such complex liquids is not firmly established and might depend strongly on the constraints used.

\section{Continuous Random Networks (CRN)}

Even though MOFs are chemically more complex than inorganic phases, the microscopic modeling of amorphous MOFs can draw inspiration from models of long-studied disordered materials, such as silica. The structure of amorphous silica has been consensually modeled as a Continuous Random Network (CRN), and there is a large body of work on this topic.\cite{Tucker2005} Random networks possess a significant degree of local order, while allowing sufficient bond distortions to have some freedom in the medium-range order, and (as desired) exhibit no long-range order. \cite{Elliott1984} If they have no broken bonds in reference to their ideal connectivity, they are said to be continuous. CRNs can be generated with a variety of approaches and possess different structures which can be validated by looking at the agreement of computed properties with experimental data: the most commonly available data is the total pair distribution function, or radial distribution functions (RDF). Just like perfect crystal\revadd{s}, CRNs should be considered as idealized, yet particularly insightful, structure\revadd{s} of the real-world materials they represent. \cite{Wright2013} 

\begin{figure}[hbtp]
\centering
\includegraphics[width=1.0 \textwidth]{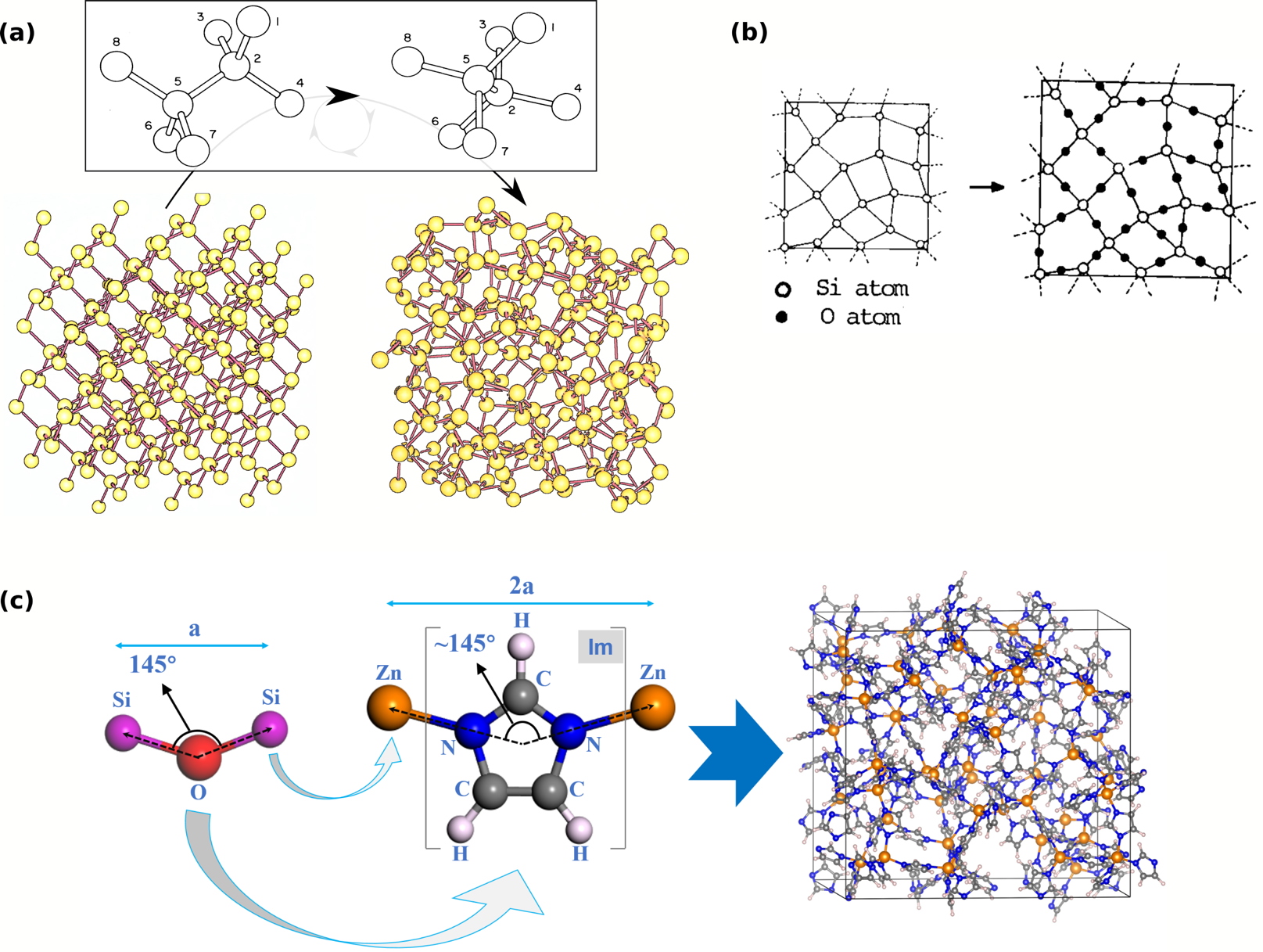}
\caption{(a) Creation of an $a$-Si continuous random network (right) following the WWW procedure, starting from a crystal (left) and performing successive bond rearrangements (inset). \cite{Wooten1987}
(b) Expansion of the $a$-Si CRN into an $a$-\ce{SiO_2} model. \cite{Ching1980}
(c) Expansion of the $a$-\ce{SiO_2} model into a model of $a$-ZIF, where the Si--O--Si linkage is replaced by Zn--Im--Zn. \cite{Wang2019}
Credit: adapted with permission from ref. ~\citenum{Wooten1987}, Copyright 1987 Academic Press, Inc. Published by Elsevier B.V.; ref. ~\citenum{Ching1980}, 
Copyright 1980 Pergamon Press Inc. Published by Elsevier Inc.; ref. ~\citenum{Wang2019}, Copyright 2019 The American Ceramic Society. 
}
\label{fig:crn_substitution}
\end{figure}

The first CRN model of an amorphous MOF was constructed for ZIF-4, on the basis of the structural similarities between ZIF-4 -- \ce{Zn(Im)_2} -- and amorphous silica $a$-\ce{SiO_2}. \cite{Bennett2010} They both share a similar short-range order with the tetrahedral coordination of the metal ion and a node--ligand--node angle of 145\textdegree. They also display corresponding features in their RDFs, while their numerous crystalline polymorphs exhibit several shared topologies. Therefore, a CRN model of $a$-\ce{SiO_2} was adapted by substituting Si and O to Zn and Im (Im = imidazolate), respectively, as represented in \autoref{fig:crn_substitution}c. This experimentally consistent \cite{Tucker2005} $a$-\ce{SiO_2} CRN had been obtained by expanding an $a$-Si model (adding O atoms between each Si as illustrated on \autoref{fig:crn_substitution}b), which was itself produced following the Wooten--Winer--Weaire (WWW) method. This widely used procedure for tetrahedrally bonded single-component networks starts from a crystal and proceeds to minimize an energy function using a simulated annealing algorithm that performs successive steps of bond rearrangements shown on \autoref{fig:crn_substitution}a. \revadd{\cite{Wooten1985, Wooten1987}}

Used as an initial configuration for the RMC refinements mentioned in \revadd{the} previous section \cite{Bennett2010}, this first CRN paved the way for CRNs to be considered as representative of $a$-ZIFs atomic structure. A similar construction procedure starting from a different $a$-Si CRN was used for the creation of an $a$-ZIF-4 model, which was studied after a \revadd{density functional theory (DFT)}-based geometry optimization. \cite{Adhikari2016} The $a$-Si model was constructed following a method predating WWW that starts from an initial centered cubic system with quasirandom coordination before also minimizing the energy, albeit with alternative bond rearrangement and minimization procedures. \cite{Guttman1981} Although this method was eclipsed by WWW which was more successful at generating large systems of arbitrary size \cite{Wooten1985}, it has the virtue of generating fully coordinated CRNs which possess no memory of the initial crystalline phase. The resulting $a$-ZIF model was validated by comparing its RDF with experimental data, and was sufficiently large (918 atoms) to realistically calculate the electronic structure, interatomic bonding and optical properties. Unlike other amorphous ZIF models, it is fully coordinated by design and is of lower density than the crystal. 
This surprisingly low density is caused by the enlargement of the $a$-Si cell before the substitution procedure as displayed on \autoref{fig:crn_substitution}c, where an arbitrary - and seemingly excessive - factor of 2 was applied.
The elastic behavior of this model was studied in a later work by successive steps of isotropic deformation of the cell and geometry optimization. The mechanical properties were computed, as well as the evolution of the electronic and optical properties under compression. \cite{Adhikari2018} An insulator-to-metal transition leading to a novel phase is claimed to be observed under extreme compression, but the apparent breaking down of the organic ligands questions the physical validity of these simulations.

This $a$-ZIF CRN model can easily be expanded to other materials of the ZIF family, by substituting the organic ligands and metal nodes to investigate their influence on derived physical properties. In particular the change in electronic and optical properties were studied for ZIF-4 and MAF-7 (featuring 1,2,4-triazole ligands) with alternating Li/B metal nodes \cite{Wang2019}, ZIF-62 (imidazolate and benzimidazolate ligands) for various ligand ratios, \cite{Xiong2020} and ZIF-4 with halogenated imidazolates (H atoms substituted with Cl or Br) \cite{Xiong2020b}. When available (i.e. only for one ZIF-62 system with a given ratio), the system was validated by comparing the RDF with experimental data, but this could not systematically be performed. The physical realism of the CRN models obtained is therefore not systematically demonstrated.

\section{Relying on crystalline models}

Because of the difficulties of directly modeling amorphous phases, some research groups have tried to address this issue by \emph{ad hoc} adaptation of modeling strategies designed for crystalline states. One such example is the creation of disordered models of UiO-66, MIL-140B and MIL-140C, three zirconium-based MOFs, by incorporating defects into the crystalline model through several possible pathways. \cite{Bennett2016} Those pathways, initially devised based on chemical intuition, were selected to display a small enough energy penalty to be thermodynamically accessible and lead to an experimentally plausible change of lattice constants. NMR (Nuclear magnetic resonance) chemical shift calculations of the defective structures derived from each pathway were performed based on \revadd{DFT}, and the NMR spectra were compared to experimental data. As several pathways led to a better agreement than the perfect crystalline model, it was inferred that some of these defects are present in the amorphous structure.

Another possible strategy, which aims not only at highlighting the presence of defects but rather at generating amorphous models, consists in simulating the phase transition from a crystalline MOF using classical molecular dynamics. Molecular dynamics (MD) reproduces the time evolution of molecules and materials by numerically integrating Newton's equations of motion, based on the knowledge of atomic interactions at a given level of theory, and can be performed at varying conditions of temperature and mechanical constraints. It is therefore a way to mimic \emph{in silico} the experimental formation routes, such as melt-quenching, by imposing adequate thermodynamic variables such as pressure or temperature. Not only does this approach generate amorphous systems, it also provides detailed knowledge of the underlying microscopic mechanism causing amorphization.

The most commonly used type of MD simulations for crystalline MOF phases is classical MD, where the interatomic potential is evaluated as an analytical function of the atomic positions, called the force field. That force field is optimized to reproduce the structure and dynamics of the framework, based on experimental data or quantum chemistry calculations. Although classical MD simulations are routinely used for crystalline MOFs,\cite{Coupry2016, Bureekaew2013} they cannot describe changes in the electronic state of the atoms. In particular, they are unable to simulate bond breaking or reformation and thus cannot simulate reconstructive phase transitions (i.e. crystal-to-crystal or crystal-to-amorphous processes in which the coordination changes).

\begin{figure}[ht]
\centering
\includegraphics[width=0.8\textwidth]{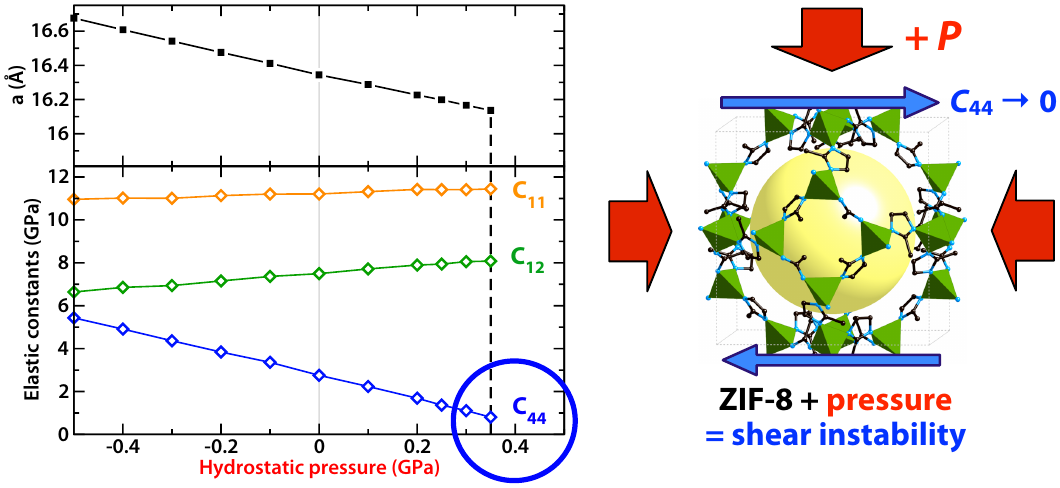}
\caption{Evolution of the unit cell edge length and elastic constants of ZIF-8 under hydrostatic pressure, from classical molecular dynamics simulations, showing the onset of amorphization at $P = 0.35$~GPa, where the shear modulus $C_{44}$ goes to zero. Adapted with permission from Ref.~\citenum{Ortiz2013}. Copyright 2013 American Chemical Society.}
\label{classical_collapse}
\end{figure}

However, some authors have attempted to use such classical MD models to study the amorphization of MOFs. For example, Ortiz et al. showed that while the classical force field could not be relied upon to describe the nature of the amorphous phase, it could still be useful to study the mechanical stability of the crystal before amorphization, and determine the onset of the phase transition. \cite{Ortiz2013} Ortiz performed the first computational study on the pressure-induced amorphization (PIA) of ZIF-8, by simulating the crystal at various pressures in a constant-stress ensemble $(N, \sigma, T)$, which allowed to estimate the amorphization pressure by identifying the point at which the phase transition occurs, illustrated on \autoref{classical_collapse}. Increasing the pressure above this threshold numerically leads to a new phase, the physical reality of which is not clear: it can be seen as a hypothetical amorphous ZIF-8 system under the constraint that no bond breaking takes place during amorphization. The same approach was also used in two separate studies, in one for illustrative purposes only \cite{Bennett2015} and in the other as a validation experiment of a newly developed force field \cite{Weng2019}. In the latter, the final amorphous system was validated by comparing the structure factor to experimental data. However, the underlying assumption of the non-reconstructive nature of this pressure-induced amorphization has yet to be directly confirmed, e.g. by \emph{in situ} measurements. It is in stark contrast with other studies that found ZIF-8 amorphization by ball milling and melt-quench\revadd{ing} to be reconstructive \cite{Cao2012, Gaillac2020}.

Finally, Bhogra et al. developed another approach to gain insight into the amorphization process by studying the instability of the crystalline phase. \cite{Bhogra2021} They combined geometry optimization calculations based on DFT and phonon-spectral analysis to investigate the amorphization pathways of MOF-5, a MOF comprised of \ce{ZnO_4} tetrahedra bonded with benzene dicarboxylate ligands. Starting with a strained crystal, they proceed by finding its unstable phonon modes (if any) to distort the structure in order to have the atomic displacements forming linear combinations of these modes, before performing a final energy minimization at fixed cell shape and volume. At sufficient strain, this last step leads to amorphization which manifests itself as an internal structural rearrangement. Although unique in the study of amorphous MOFs and performed in the zero-Kelvin limit, this approach has the benefit of hinting at the microscopic mechanism for pressure-induced amorphization --- even though pressure and temperature are not explicitly present.

\section{\emph{Ab Initio} Molecular Dynamics}

Unlike classical MD, \emph{ab initio} molecular dynamics (AIMD) is a method that allows a full description of the electronic state of the system at the quantum chemical level and can describe the formation and breaking of chemical bonds. It is therefore highly suited for the modeling of reconstructive phase transitions. AIMD, of which there are several ``flavors'', combines the modeling of the equations of motion of the nuclei and the quantum nature of the electrons. In its most common flavor, the Born--Oppenheimer dynamics, the Schr\"{o}dinger equation is solved at each MD time step, typically using Density Functional Theory (DFT). More computationally demanding than classical MD, AIMD can only simulate phenomena at smaller time and length scales. However, using modern high-performance computing resources, it is tractable for MOFs with a few hundred atoms in the unit cell, such as most ZIFs, for times ranging from tens to hundreds of picoseconds.

\begin{figure}[ht]
\centering
\includegraphics[width=\textwidth]{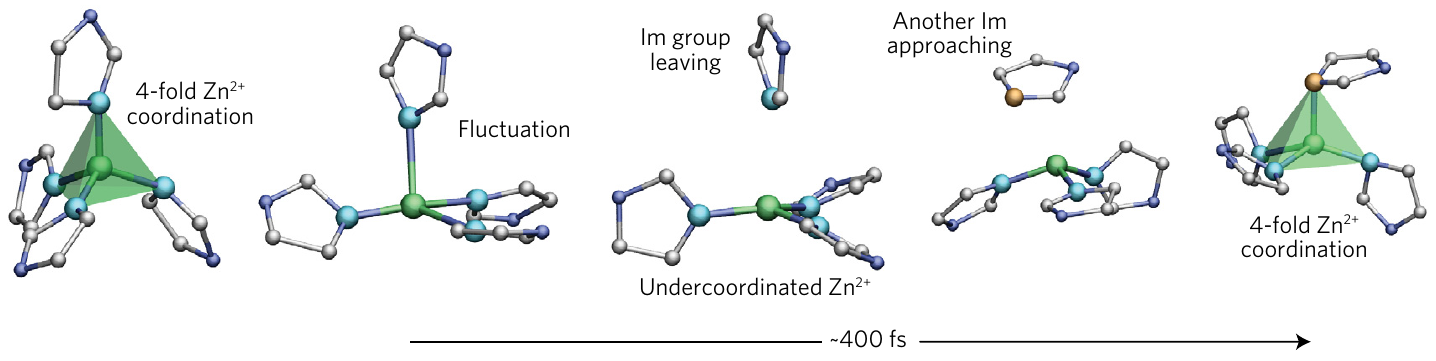}
\caption{Visualization of a representative imidazolate (Im) exchange event in the microscopic mechanism of ZIF-4 melting, as observed by \emph{ab initio} molecular dynamics. Reproduced from Ref.~\citenum{Gaillac2017}. Copyright 2017 Nature Publishing Group.}
\label{aimd_mechanism}
\end{figure}

The first AIMD simulations in the field of amorphous MOFs studied the melting and subsequent amorphization of ZIF-4 by \revadd{quenching}. \cite{Gaillac2017} Starting from a single unit cell of the ZIF-4 crystal, several distinct simulations were performed in the constant-volume $(N, V, T)$ ensemble with temperatures ranging from 300 to 2250~K, and adequate volume to reproduce the experimentally measured densities. This procedure did not aim at generating a glass model, but rather at identifying the melting transition and studying the microscopic mechanism involved. By presenting a representative imidazolate exchange event at the origin of bond rearrangements, \autoref{aimd_mechanism} exemplifies the atomistic insight permitted by AIMD, looking at \revadd{structural and dynamical} details. Additionally, AIMD was used to investigate the thermodynamics of melting and characterize the generated liquid ZIF, in synergy (and good agreement) with \emph{in situ} variable temperature X-ray and \emph{ex situ} neutron pair distribution function experiments performed in the same study. The same procedure was subsequently performed on two other networks with different topologies: ZIF-zni, chemically identical, and ZIF-8, composed of different ligands. \revadd{\cite{Gaillac2018}} As the details of the DFT methodology are largely independent from the topology, the chemical similarity of the frameworks ensure\revadd{d} a smooth transferability. \revadd{By computing the melting temperature of each framework and comparing it to its experimentally determined decomposition temperature, this second work helped explain why some frameworks are experimentally observed to melt while others collapse before melting.}

While particularly insightful into the melting dynamics, these two studies did not explicitly generate atomistic models of ZIF glasses. This was later completed by a \revadd{subsequent} work simulating the entire glass formation procedure, through melting and quenching\revadd{, which highlighted the distinct structural properties of the glasses compared to both the liquid and crystalline phases.} \cite{Gaillac2020} Three ZIFs crystals -- ZIF-4, ZIF-8 and SALEM-2 (which features imidazolate ligands like ZIF-4, but has the same topology as ZIF-8) --- were first melted at 1500~K, before being quenched to 300~K with a \revadd{cooling} rate of 50~K/ps. Because the high computational cost of AIMD puts strong limits on the system size, and makes it unfeasible in this case to study larger supercells, 10 quenching simulations were performed for each material to reduce the impact of finite size effects and get a statistically representative description of the glasses. We should note, however, that the computational cost also limits the total simulation time, and means that the \revadd{cooling} rates attained are several orders of magnitude higher than any achievable in the laboratory. \cite{Elliott1984} We note that although this is an inherent limitation of any molecular dynamics simulation, it is more pronounced for \emph{ab initio}.

The use of the constant-pressure MD simulation in the $(N, P, T)$ ensemble would prove a potential improvement of these works, as it would allow the joint study of both dynamical and thermal effects without having to input the system density. However, \emph{ab initio} simulations of such systems are notoriously challenging to equilibrate, as these frameworks respond very sensitively to small external stimuli and thus to small deviations in the computation of the stress tensor. \cite{Haigis2014} As of today, no computational scheme has been demonstrated to accurately describe the equilibrium between the three phases involved (crystal, liquid and glass) and the respective values of their density.

However, AIMD was used to help study the effect of pressure on various types of amorphization processes, through the application of high pressure, high temperature, or both. Widmer et al. studied the low-temperature melting of two ZIFs, ZIF-4 and ZIF-62, under the application of hydrostatic pressure. \cite{Widmer2019} Multiple simulations were performed in the constant-pressure $(N, P, T)$ ensemble around the thermal amorphization temperature with various pressures in the 0.1 to 5~GPa range. Similarly to the aforementioned melting studies \cite{Gaillac2017} \cite{Gaillac2018} this work gave insight into the melting process under pressure, but did not aim at providing an atomistic model of the final, fully equilibrated molten frameworks.

Finally, two works tackled pressure-induced amorphization by running AIMD simulations with increasing pressure, going above the amorphization onset, with a different methodology. Erkartal et al. developed an \emph{ad hoc} technique that proceeds by running successive short out-of-equilibrium isoenthalpic-isobaric $(N, P, H)$ MD runs at various pressures.  Neither the system enthalpy nor temperature are actually controlled in the process, making it somewhat akin to an energy minimization algorithm. 
The first work studied the MOF-5 amorphization under hydrostatic pressure by investigating the change in structural and electronic properties and provided the first amorphous MOF-5 atomistic model ever reported. \cite{Erkartal2018} The second one investigated pressure-induced amorphization of ZIF-8 both under hydrostatic and uniaxial stress. The resulting amorphous ZIF-8 was obtained without bond-breaking, and the method was validated by comparing the RDF with experimental data.  \cite{Erkartal2020}

\section{Reactive force fields}

Despite its chemical accuracy and ability to simulate reconstructive phase transitions, AIMD significant computational cost limits its use to small systems on rather short time scales (hundreds of ps) and all but prohibits its use for high-throughput screening. Reactive force fields are empirical force fields that possess connection-dependent terms, enabling the simulation of bond breaking and reformation. Compared to \emph{ab initio} methods, they trade accuracy for lower computation cost. \cite{Senftle2016}

Their use to study amorphous MOFs is so far \revadd{primarily focused on melt-quenched} ZIF glasses. The systems were simulated using ReaxFF, a flavor of reactive force fields which interatomic potentials are functions of bond-order, itself calculated from interatomic distances. This was made possible by the development of a reactive force field conceived for the study of Zn-Imidazolates complexes in aqueous, validated for these systems with \emph{ab initio} data. \cite{Shin2021} This ReaxFF force field was then used to generate amorphous ZIFs, after a limited preliminary validation on crystalline structures. \cite{Yang2018} As this force field was neither originally developed --- nor subsequently shown --- to accurately reproduce the geometry of the \ce{Zn(Im)_4} tetrahedra, key to the framework properties, predictions made through its use would benefit from further validation against \emph{ab initio} data.

\begin{figure}[ht]
\centering
\includegraphics[width=0.7 \textwidth]{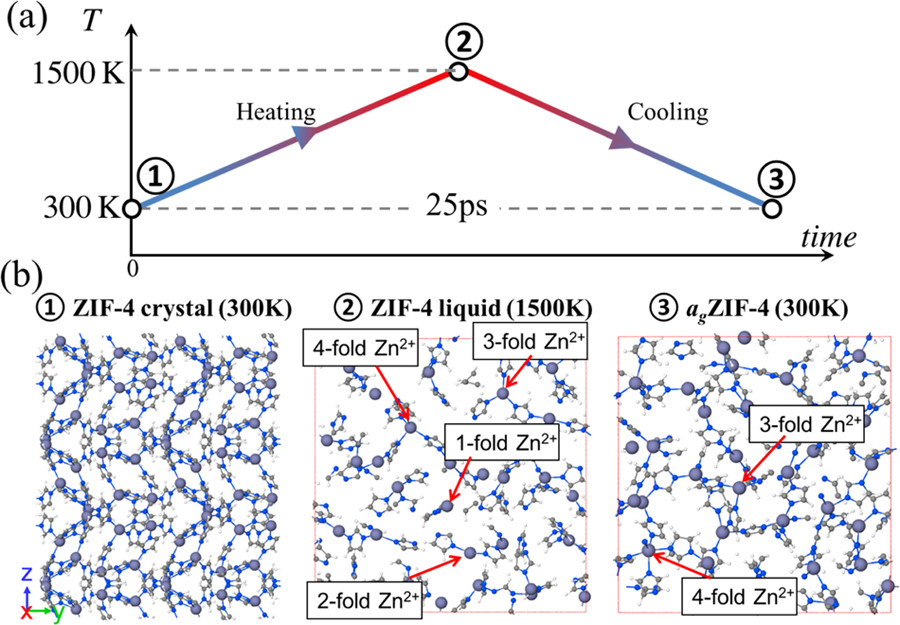}
\caption{Representation of the melt-quench\revadd{ing} process applied to ZIF-4 using the ReaxFF reactive force field. (a) Evolution of temperature over time. (b) Successive snapshots at three different times, in the crystal, liquid, then glass state\revadd{s}. 
Reprinted with permission from Ref.~\citenum{Yang2018}. Copyright 2018 American Chemical Society.}
\label{reaxff_meltquench}
\end{figure}

In a first work, three ZIF crystals, ZIF-4, ZIF-62 and ZIF-77 (2-nitroimidazolate ligands), were heated above melting temperature before being quenched in the constant-pressure $(N, P, T)$ ensemble as illustrated on \autoref{reaxff_meltquench}. The \revadd{heating/cooling} rates were of 96~K/ps, higher than the AIMD melt-quench\revadd{ing} work \cite{Gaillac2020}. The inexpensiveness of the method allowed to simulate a \revadd{$(2\times2\times2)$} supercell, limiting the finite size effects. The approach was validated for ZIF-4 by comparing the RDF to experimental data and several properties to the first \emph{ab initio} work on $a$-ZIF-4 \cite{Gaillac2017}.  Compared to AIMD, the ReaxFF simulation yielded the same heat capacity and reproduced the same relationship, albeit more pronounced, between temperature and the undercoordination of Zn nodes. \revadd{However, unlike ZIF-4 glasses later obtained by an AIMD melt-quenching simulation \cite{Gaillac2020} which preserved the density (1.2) and porosity of the crystal, the ReaxFF glass was of significantly higher density (1.6) and lost its porosity.}

This melt-quench\revadd{ing} procedure \revadd{with ReaxFF} was then replicated on various ZIFs in further works, albeit with a slower \revadd{heating/cooling} rate of 24~K/ps, to study properties that would otherwise be prohibitively expensive with AIMD. An amorphous ZIF-62 model was generated to study the fracture toughness of the glass by inducing a precrack in an enlarged \revadd{$(2\times6\times4)$} supercell  before a stepwise elongation. \cite{To2020} It was validated by comparing to experimental data the RDF and multiple mechanical properties, i.e. Young modulus, Poisson's ratio and fracture toughness. A subsequent study performed equivalent simulations for ZIF-4 and ZIF-76 (mix of imidazolate and 5-methylbenzimidazolate ligands), before investigating the structural origin of the facture behavior. \cite{To2021} 
\revadd{Yet a}nother work computed the thermal conductivity of amorphous ZIF-4, ZIF-62 and ZIF-8, which were validated by comparison with experimental values. \cite{Sorensen2020} The glass was shown to have a higher conductivity than the crystal, a very unusual relation, and the atomistic models were key to investigate the structural origin of this phenomenon.
\revadd{Finally, the obtained atomistic models were used, along with experimental data, to develop a topological constraint model which predicted the glass transition temperatures of any ZIF sharing the topology of ZIF-4 (cag) with an arbitrary mix of three ligands (imidazolate, benzimidazolate and 5-methylbenzimidazolate). \cite{Yang2018b}}

\revadd{In addition to this large body of work on melt-quenched ZIFs glasses, an earlier ReaxFF-based force field, initially developed for the interaction of glycine with a copper surface and not validated for MOFs, was used to study the mechanical amorphization of three copper based MOFs. \cite{Mohamed2021}
These frameworks, chosen to contain non-accessible regions in their crystalline form, were subjected to shear or compressive deformations performed in the constant-pressure $(N, P, T)$ ensemble. At moderate strain levels, they displayed an enhancement of their porosity, made possible by the breakage of metal--linker bonds which lead to partial amorphization.
}

\section{Polymerization algorithms}

\begin{figure}[hbtp]
\centering
\includegraphics[width=1.0 \textwidth]{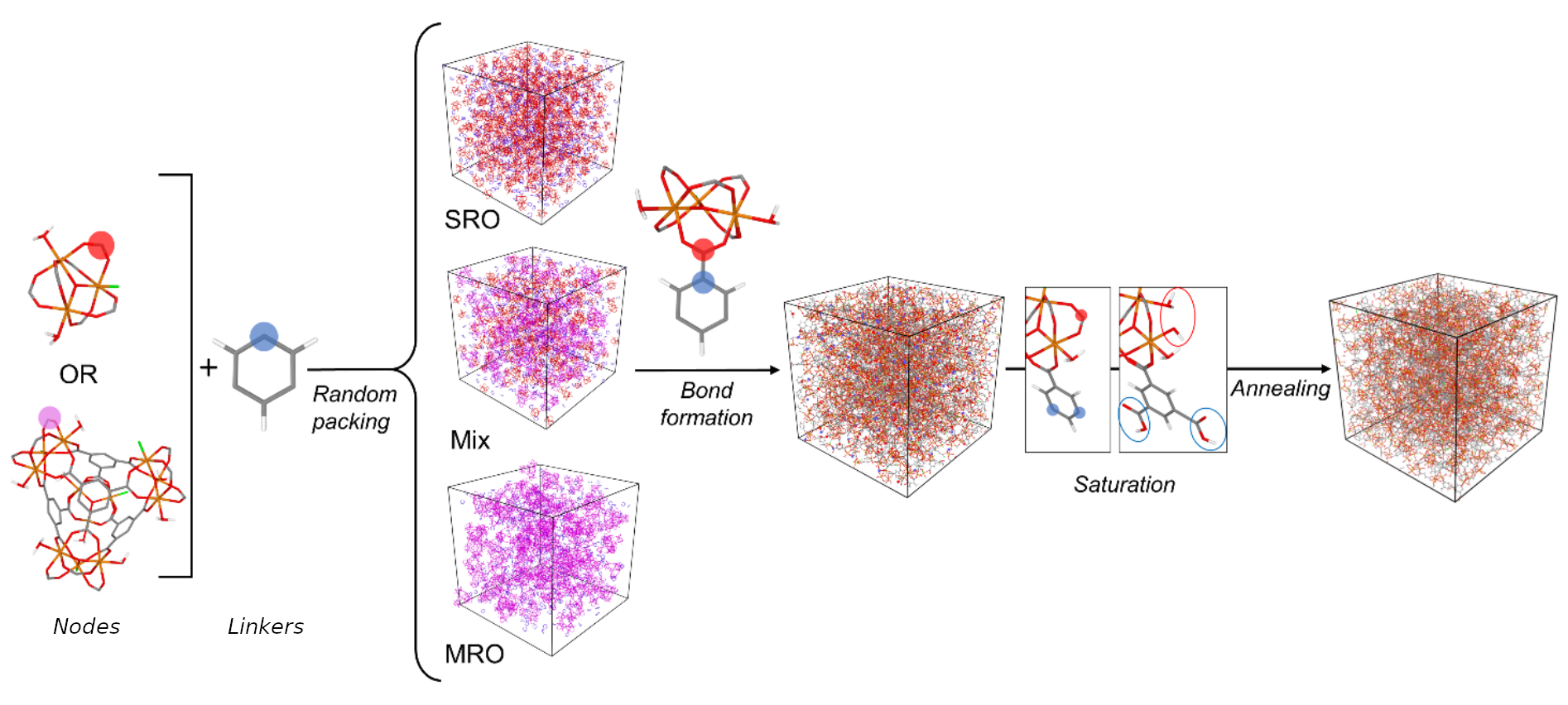}
\caption{Representation of the procedure used to build three amorphous models (SRO, MIX and MRO, see text for details) of Fe-BTC with the Polymatic method.
Adapted from Ref.~\citenum{Sapnik2021}, under the terms of the Creative Commons CC BY license.}
\label{polymatic_procedure}
\end{figure}

Although classical MD cannot be used to simulate a reconstructive phase transition, it can still play a role in the generation of atomistic models using simulated assembly/polymerization-based modeling. Polymatic, a generalized simulated polymerization algorithm, was first developed for amorphous polymers and later used to generate amorphous MOFs following a well-defined procedure illustrated on \autoref{polymatic_procedure}. Starting from a low-density random packing of the building units of the amorphous material, i.e. the metal nodes and organic linkers for MOFs, successive steps of bond formation and structural rearrangement are performed. Bond formation is only allowed between predefined reactive sites and considered when two sites are within a defined cut-off distance. Structural rearrangement at turns involves energy minimizations and MD steps in the $(N, V, T)$ or $(N, P, T)$ ensemble. When no further bonds can be formed, the structure is annealed using a multistep MD protocol that applies for a limited time artificially high pressure to compress the system to a reasonable density. \cite{Abbott2013} As the bonds are at all time well-defined, classical force fields can and are employed for energy minimizations and MD runs.

The first amorphous MOF generated with Polymatic is the well-studied ZIF-4, of which 5 atomistic models were generated to sample different regions of configuration space. \cite{Thornton2016} The validation of the approach was limited to the comparison of the density and pore volume fraction to experimental values.

A more comprehensive study by Sapnik et al. modeled for the first time two MOFs, Fe-BTC and Basolite F300, one of its commercial forms. \cite{Sapnik2021} As illustrated on \autoref{polymatic_procedure}, they contain two sorts of nodes, trimer units (\ce{FeO_6} octahedra that cluster around a shared oxo-anion) and tetrahedral assemblies (4 assembled trimers via an organic linker), and 1,3,5-benzenetricarboxylate anions as linkers. Before this work, \revadd{their} atomic-scale structure\revadd{s were} unknown, and \revadd{they were} previously being described as either disordered, amorphous or nanocrystalline. \emph{Ab initio} methods are not an option due to \revadd{the considerable number of atoms (more than 10,000) needed to describe such systems}, and no reactive force fields are available for \revadd{their} chemical composition\revadd{s}. UFF4MOF, the extension of the Universal Force Field parametrized for the description of MOFs \cite{Coupry2016}, was used in this study after validation on trimers and linkers with DFT. As the node ratio was unknown, three models were built: a short-range order model (SRO) containing {100\%} trimers, a mixed model (MIX) containing a 50/50\% mixture of trimers and tetrahedra, and a medium-range order model (MRO) containing {100\%} tetrahedra. Each model was built 5 times and properties were averaged. Guided by RDF similarity and pore analysis, the authors concluded that two of these models could be considered representative of the atomic structure of Fe-BTC and Basolite F300. From this they deduced a structure--property relationship between the degree of tetrahedral assembly and porosity. This work \revadd{on two carboxylate MOFs} exemplifies how the comparatively low computational cost of this approach, along with the versatility of classical force fields, make it viable for the generation and investigation of various disordered structures\revadd{, including outside of the well-studied ZIF family.}

\section{Conclusions and perspectives}

There is an increasing interest in the amorphous states of metal-organic frameworks (MOFs) and porous coordination polymers, but the determination of their framework structure\revadd{s} at the microscopic scale is difficult. We reviewed and compared the existing methodologies available for amorphous MOFs, based on both experimental data and \revadd{numerical} simulation\revadd{s}, which are Reverse Monte Carlo (RMC) methods, Continuous Random Networks (CRN), classical and \emph{ab initio} molecular dynamics, reactive force fields, and polymerization algorithms. From that review, it is clear that in most cases, the problem of modeling is mathematically underdetermined, and it is difficult to arbitrate between different possible models. This is due to the relative lack of available experimental data, with only indirect structural information obtained from diffraction measurements. 

In order to improve the quality of amorphous models in the future, we see the need for wider studies, integrating many different experimental techniques, in order to provide \emph{in situ} data, for example by spectroscopic methods: infrared and Raman, NMR, etc.\cite{Madsen2020} In addition to providing direct insight into the nature of the amorphous phases of MOFs, such data could be used as a benchmark to test the different types of microscopic models generated.
\revadd{Databases of amorphous porous materials could then be expanded, along the lines of what has been done for years for the crystalline phases, and help accelerate efforts to model these systems \cite{Thyagarajan2020}.
}

We also find that there is, in the existing literature, a lack of direct and in-depth comparison of the models. In particular, their geometrical, physical and chemical properties have not been systematically compared against each other. While the generating methods vary a lot (some are purely mathematical, some rely on physical or chemical insight, some perform direct molecular simulations), the models produced have very different characteristics that should be systematically computed and contrasted: density, porosity, framework coordination, topology, etc.

Finally, it appears to us that no single modeling method can currently yield an accurate microscopic representation of the MOF glasses, which suggest\revadd{s} the development of multi-scale modeling strategies, combining the strength of the different methods already available. This would allow bridging the simulations performed at various scales: geometric, classical, quantum. One area of active development that appears very promising is the development of machine learnt (ML) potentials for the description of atomic interactions. \cite{Behler2017, Eckhoff2019} This could lead to a new generation of specific and accurate reactive potentials for the description, in particular, of the coordination interactions that are key to the metal--ligand bond breaking and formation during amorphization and in the glassy state.

\begin{acknowledgement}
The authors thank Kim Jelfs and Tom Bennett for discussions and ongoing collaboration on this topic.
\end{acknowledgement}

\section*{Biographies}

\includegraphics[width=0.2 \linewidth]{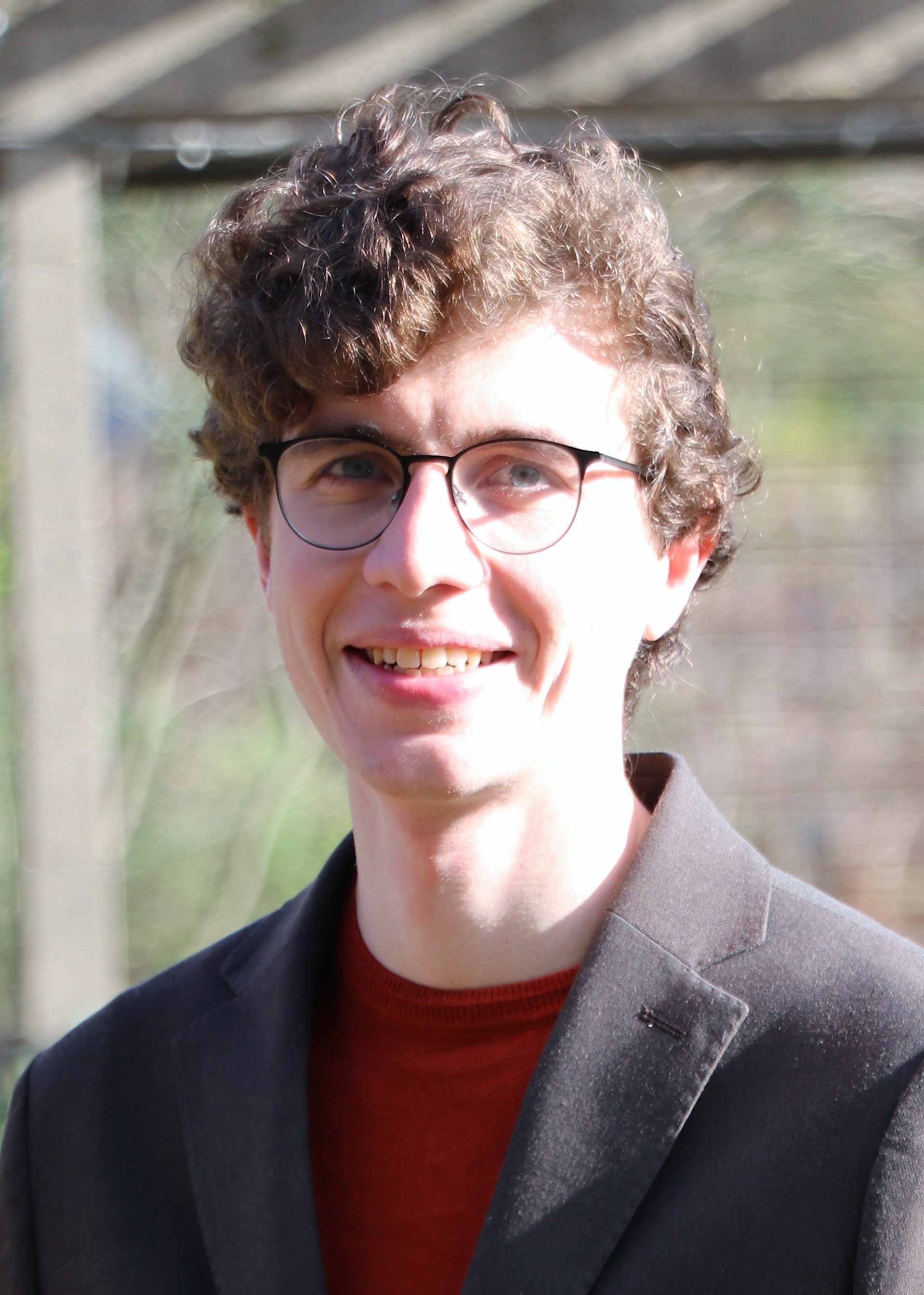}

\noindent\textbf{Nicolas Castel} is currently a PhD candidate at PSL University (France). He previously received a MASt in Physics (Part III) from the University of Cambridge (UK) in 2019 and an Engineering Degree from \'{E}cole Polytechnique (France) in 2018.
In his PhD project, he is using multiple molecular simulation techniques to study amorphous metal-organic frameworks relevant for their application in \ce{CO_2} separation.

\noindent\includegraphics[width=0.2 \linewidth]{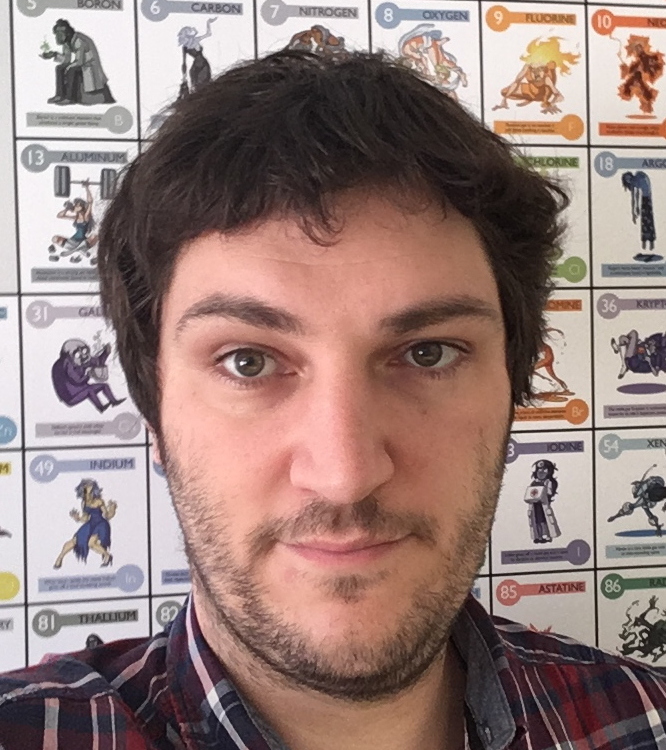}

\noindent\textbf{Fran\c{c}ois-Xavier Coudert} is a senior researcher at the French National Centre for Scientific Research (CNRS) and a professor at PSL University / \'{E}cole normale sup\'{e}rieure. His group applies computational chemistry methods at various scales to investigate the physical and chemical properties of nanoporous materials, their responses to external stimuli, and the behavior of fluids at fluid/solid interfaces. He obtained his Ph.D. from the Universit\'{e} Paris-Sud (France) in 2007 for work on the properties of water and solvated electrons confined in zeolite nanopores. He worked as a postdoctoral researcher at University College London (UK) on the growth of metal-organic frameworks on surfaces, before joining CNRS in 2008.

\bibliography{article}

\begin{tocentry}
\includegraphics[width=\linewidth, clip, trim=0 710 0 0]{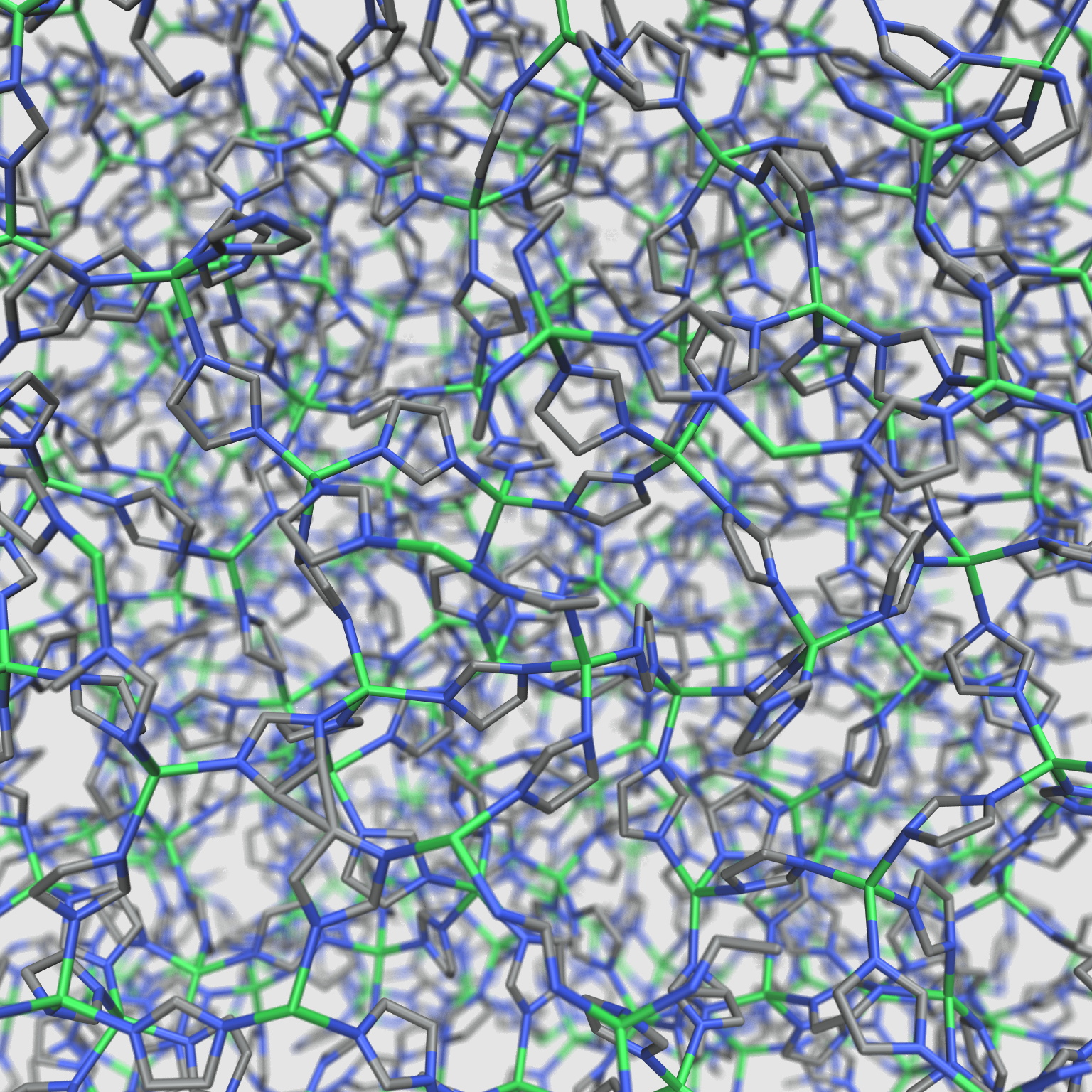}
\end{tocentry}

\end{document}